\documentclass{pasj01}
\usepackage{url}

\begin{document}
\SetRunningHead{Tanaka et al.}{Deep HSC Imaging of COSMOS}

\title{Deep Optical Imaging of the COSMOS Field with Hyper Suprime-Cam Using Data from the Subaru Strategic Program and the University of Hawaii}

\author{
  Masayuki Tanaka\altaffilmark{1},
  G\"{u}nther Hasinger\altaffilmark{2},
  John D. Silverman\altaffilmark{3},
  Steven Bickerton\altaffilmark{4},
  Hisanori Furusawa\altaffilmark{1},
  Yuichi Harikane\altaffilmark{5,6},
  Esther Hu\altaffilmark{2},
  Hiroyuki Ikeda\altaffilmark{1},
  Yanxia Li\altaffilmark{2},
  Henry J. McCracken\altaffilmark{7},
  Paul A. Price\altaffilmark{8},
  Michael A. Strauss\altaffilmark{8},
  Michitaro Koike\altaffilmark{1},
  Yutaka Komiyama\altaffilmark{1,9},
  Sogo Mineo\altaffilmark{1},
  Satoshi Miyazaki\altaffilmark{1,9},
  Atsushi J. Nishizawa\altaffilmark{10},
  Tadafumi Takata\altaffilmark{1,9},
  Yousuke Utsumi\altaffilmark{11},
  Yoshihiko Yamada\altaffilmark{1},
  Naoki Yasuda\altaffilmark{3}
}
\altaffiltext{1}{National Astronomical Observatory of Japan, 2-21-1 Osawa, Mitaka, Tokyo 181-8588, Japan}
\altaffiltext{2}{Institute for Astronomy, University of Hawaii, 2680 Woodlawn Drive, Honolulu, HI 96822, USA}
\altaffiltext{3}{Kavli Institute for the Physics and Mathematics of the Universe (Kavli IPMU, WPI), University of Tokyo, Chiba 277-8582, Japan}
\altaffiltext{4}{Orbital Insight, 100 W. Evelyn Ave. Mountain View, CA 94041}
\altaffiltext{5}{Institute for Cosmic Ray Research, The University of Tokyo, 5-1-5 Kashiwanoha, Kashiwa, Chiba 277-8582, Japan}
\altaffiltext{6}{Department of Astronomy, Graduate School of Science, The University of Tokyo, 7-3-1 Hongo, Bunkyo, Tokyo, 113-0033, Japan}
\altaffiltext{7}{Institut d'Astrophysique de Paris, CNRS \& UPMC, UMR 7095, 98 bis Boulevard Arago, 75014, Paris, France}
\altaffiltext{8}{Department of Astrophysical Sciences, Princeton University, 4 Ivy Lane, Princeton, NJ 08544}
\altaffiltext{9}{Department of Astronomy, School of Science, Graduate University for Advanced Studies (SOKENDAI), 2-21-1, Osawa, Mitaka, Tokyo 181-8588, Japan}
\altaffiltext{10}{Institute for Advanced Research, Nagoya University, Nagoya 464-8602, Aichi, Japan}
\altaffiltext{11}{Hiroshima Astrophysical Science Center, Hiroshima University, 1-3-1 Kagamiyama, Higashi-Hiroshima, Hiroshima, 739-8526, Japan}

\email{masayuki.tanaka@nao.ac.jp}


\maketitle
\newcommand{\commentblue}[1]{\textcolor{blue} {\textbf{#1}}}
\newcommand{\commentred}[1]{\textcolor{red} {\textbf{#1}}}
\newcommand{\bs}[1]{ {\boldsymbol{#1}}}

\begin{abstract}
  We present the deepest optical images of the COSMOS field based on
  a joint dataset taken with Hyper Suprime-Cam (HSC) by the HSC Subaru Strategic
  Program (SSP) team and the University of Hawaii (UH).  The COSMOS field is one of the key
  extragalactic fields with a wealth of deep, multi-wavelength data.  However,
  the current optical data are not sufficiently deep to match with, e.g.,
  the UltraVista data in the near-infrared.  The SSP team and UH have joined forces to produce
  very deep optical images of the COSMOS field by combining data from both teams.
  The coadd images reach depths of $g=27.8$, $r=27.7$, $i=27.6$,
  $z=26.8$, and $y=26.2$ mag at $5\sigma$ for point sources based on flux uncertainties
  quoted by the pipeline and they cover essentially
  the entire COSMOS 2 square degree field.  The seeing
  is between 0.6 and 0.9 arcsec on the coadds.  We perform several quality checks
  and confirm that the data are of science quality; $\sim2$\%  photometry
  and 30~mas astrometry.  This accuracy is identical to the
  Public Data Release 1 from HSC-SSP.  We make the joint dataset including
  fully calibrated catalogs of detected objects available to
  the community at \url{https://hsc-release.mtk.nao.ac.jp/}.
\end{abstract}

\section{Introduction}
\label{sec:introduction}

Over the last decade,
our understanding of the evolution of galaxies, supermassive black holes and their location within large-scale structures out to high redshift ($z\sim6$) has vastly improved with the advent of both wide and deep multi-wavelength surveys (e.g., COSMOS, VVDS, AEGIS). As an example, the HST COSMOS survey has measured the changes in the structural properties of galaxies as a function of their environment \citep{scoville07,Capak2007,Tasca2009, Scoville2013}. This survey has stimulated a great deal of observations in other wavelengths.  X-ray observations with Chandra \citep{Elvis2009,Civano2016} and XMM-Newton \citep{Hasinger2007} have identified  rapidly growing supermassive black holes and have elucidated the role of mergers in triggering nuclear activity \citep{Cisternas2011,Silverman2011}. Spitzer \citep{Sanders2007} and Herschel \citep{Lutz2011} observations have provided a robust measure of the stellar mass content out to $z\sim6$ and unveiled star formation obscured at shorter wavelengths out to $z\sim4$. The COSMOS field now contains close to one million galaxies with photometric redshifts \citep{Laigle2016} and $\sim40000$ objects with spectroscopic redshifts \citep{Lilly2007,Trump2007, Lilly2009, Silverman2015}. As a result, the COSMOS field is likely to be an important reference for future wide field surveys from both the ground (LSST; \cite{ivezic08}) and space (Euclid;\cite{laureijs11}, WFIRST; \cite{spergel15}).

The backbone of any such survey is deep optical imaging from the ground. With respect to the COSMOS survey, the Subaru Telescope with Suprime-Cam \citep{miyazaki02} has provided broad, medium, and narrow band imaging reaching faint magnitudes across the full area \citep{Taniguchi2007, Taniguchi2015}. Subaru provides quality imaging across a wide field-of-view  by placing its imager at prime-focus, while maintaining high stability thus resulting in an optimal PSF across the field. However, even with Suprime-Cam, the observations of COSMOS required a mosaic of 9 pointings to cover the entire field and that introduced inhomogeneities in the data set.

Subaru's next generation optical camera, namely Hyper-Suprime-Cam (HSC; \cite{miyazaki17,komiyama17}), has been built and commissioned, and is now operating at full capability. HSC has 104 Hamamatsu red-sensitive science CCDs that simultaneously image 1.77 square degrees of the sky. Both broad ($grizy$) and narrow band filters are available \citep{kawanomoto17} and data are processed with the on-site reduction system for real-time quality assurance \citep{furusawa17}. 　With this new imaging capability, there is an intense effort to image large regions on the sky. In particular, the HSC team has a 3-layered Subaru Strategic Program (SSP; \cite{aihara17a,aihara17b}) with a total allocation of 300 nights.  The Wide layer is designed to image 1400 square degrees in all five broad bands, with a $5\,\sigma$ point-source depth of $r \approx 26$.  Mostly within the Wide area footprint, there is a Deep layer covering approximately 26 square degrees and an UltraDeep layer of about 3.5 square degrees that includes COSMOS and SXDS.  Independently from the SSP, the University of Hawaii (UH) is carrying out a 100 square-degree survey of the North Ecliptic Pole called HEROES (PI G. Hasinger).

The COSMOS field is of much interest to both communities (e.g., it is the primary photo-$z$ calibration field for SSP; \cite{tanaka17}).  It can now be imaged in one pointing with HSC, allowing us to significantly reduce the inhomogeneities in the current data, and there have been a lot of observations of COSMOS independently conducted by the SSP and UH teams.  In this paper, we present to the broader community a combined data set that merges the HSC observations of COSMOS taken by the two teams. Our aim is to describe the observations and quality of the deep data set for the five broad filter bands ($grizy$). The images and catalogs can all be retrieved from the HSC-SSP data release site\footnote{\url{https://hsc-release.mtk.nao.ac.jp}}.  Unless otherwise stated, all magnitudes are reported in the AB magnitude system.

\section{Data}
\label{sec:data}

We combine the HSC data collected by the SSP team and University of
Hawaii (UH) up through the first half of 2015.
The SSP team has observed COSMOS in both broad and narrow-band filters,
but only the broad-band data are included in the joint processing.
The broad-band SSP data included here are
the same as those used in the recent public data release 1 (PDR1; \cite{aihara17a})
and they were taken between 2014-03-25 and 2015-05-21.
The UH observations were conducted between 2014-03-26 and 2015-01-20.
The UH data do not include the $r$-band and hence the $r$-band images in the joint
data set are identical to those in the PDR1.  All of the raw data used in
the processing are publicly available through SMOKA\footnote{\url{http://smoka.nao.ac.jp/}}.

The data are processed with hscPipe \citep{bosch17}, a version of
the LSST stack \citep{ivezic08,axelrod10,juric15}, and the
astrometry and photometry are calibrated against the PanSTARRS1 (PS1) PV2
catalog \citep{schlafly12,tonry12,magnier13}. 
We use the same version of the pipeline (v4.0.2) with exactly the
same setup as used in the first data release (DR1), and thus we inherit
all the issues in DR1 in this joint data products.  
Refer to \citet{aihara17a} and data release site for the list of known issues.
The data processing is described in \citet{aihara17a} and algorithmic details are
given in \citet{bosch17}.  We only give a brief overview of the processing here.

We first detrend CCD characteristics by subtracting bias and applying dome flats.
Fringes are subtracted only from the $y$-band data as the other bands do not show fringes.
We then subtract the sky and detect sources in each CCD in order to calibrate the astrometry
and photometry.  As we have dithered around the COSMOS field, the same objects are
observed in different CCDs in different {\it visits} (i.e., exposures).  We apply the joint
calibration to refine the astrometry and photometry using multiple visits.  This calibration
is done for each {\it tract}, which is a 1.7$\times1.7$ degree pre-defined, iso-latitude tile.
A tract is divided into 9$\times$9 {\it patches}, each of which is 4,200$\times$4,200 pixels,
in order to parallelize the processing.
Coaddition of individual visits is done for each patch, and objects are detected and measured
again to generate the final source catalog in each coadd.

\begin{figure*}
  \begin{center}
    \includegraphics[width=7cm]{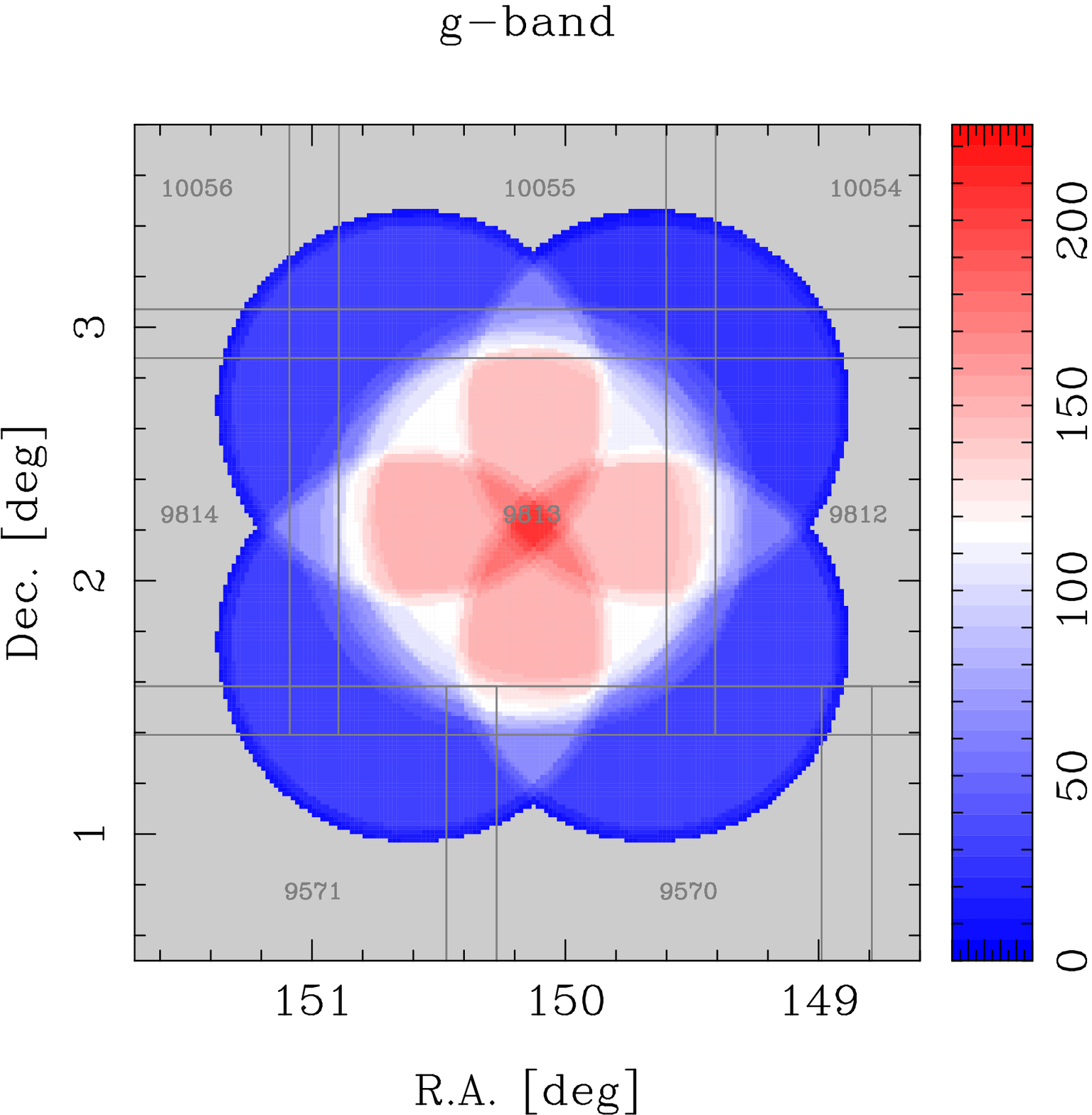}\hspace{0.5cm}
    \includegraphics[width=7cm]{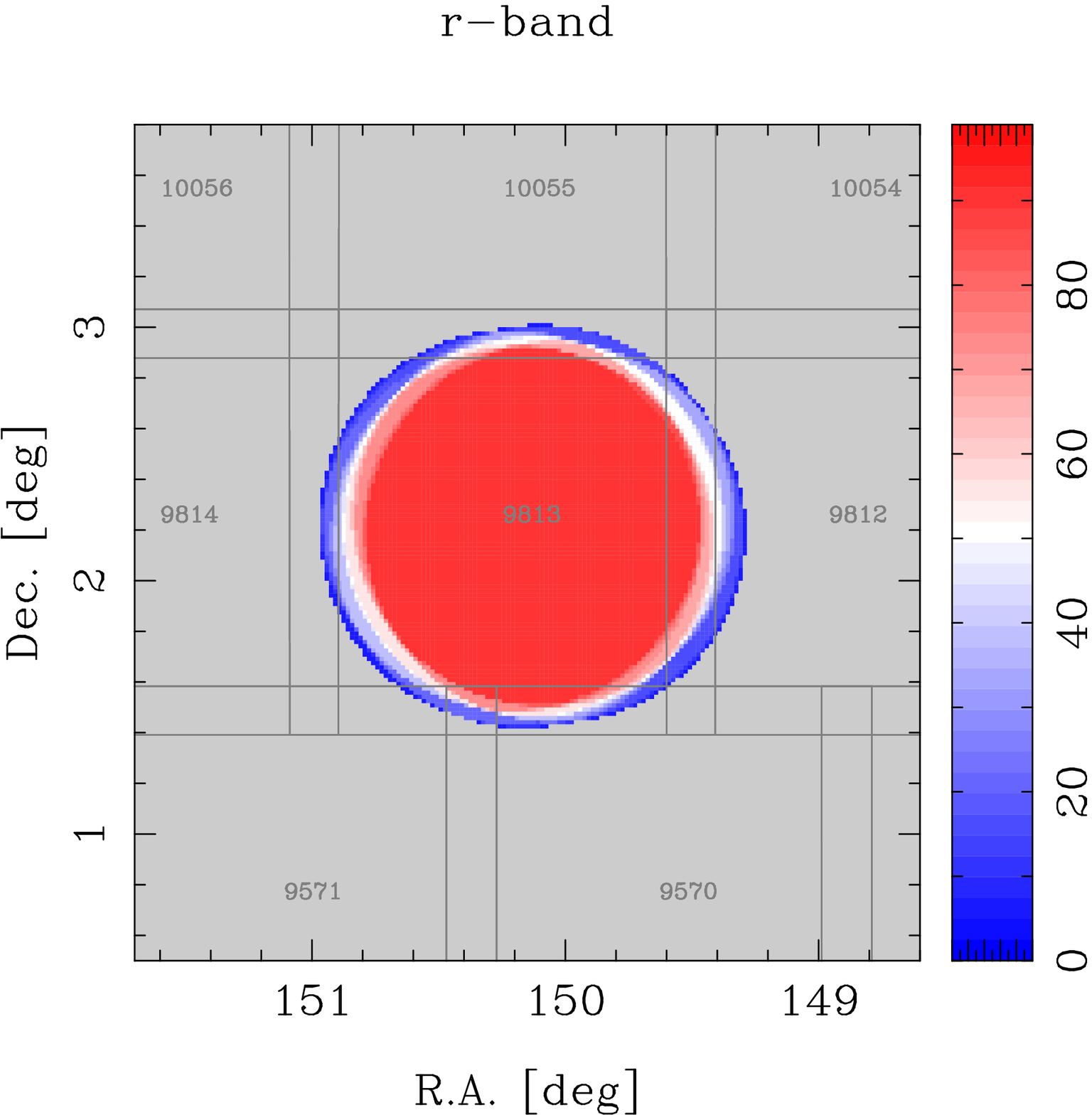}\\\vspace{0.5cm}
    \includegraphics[width=7cm]{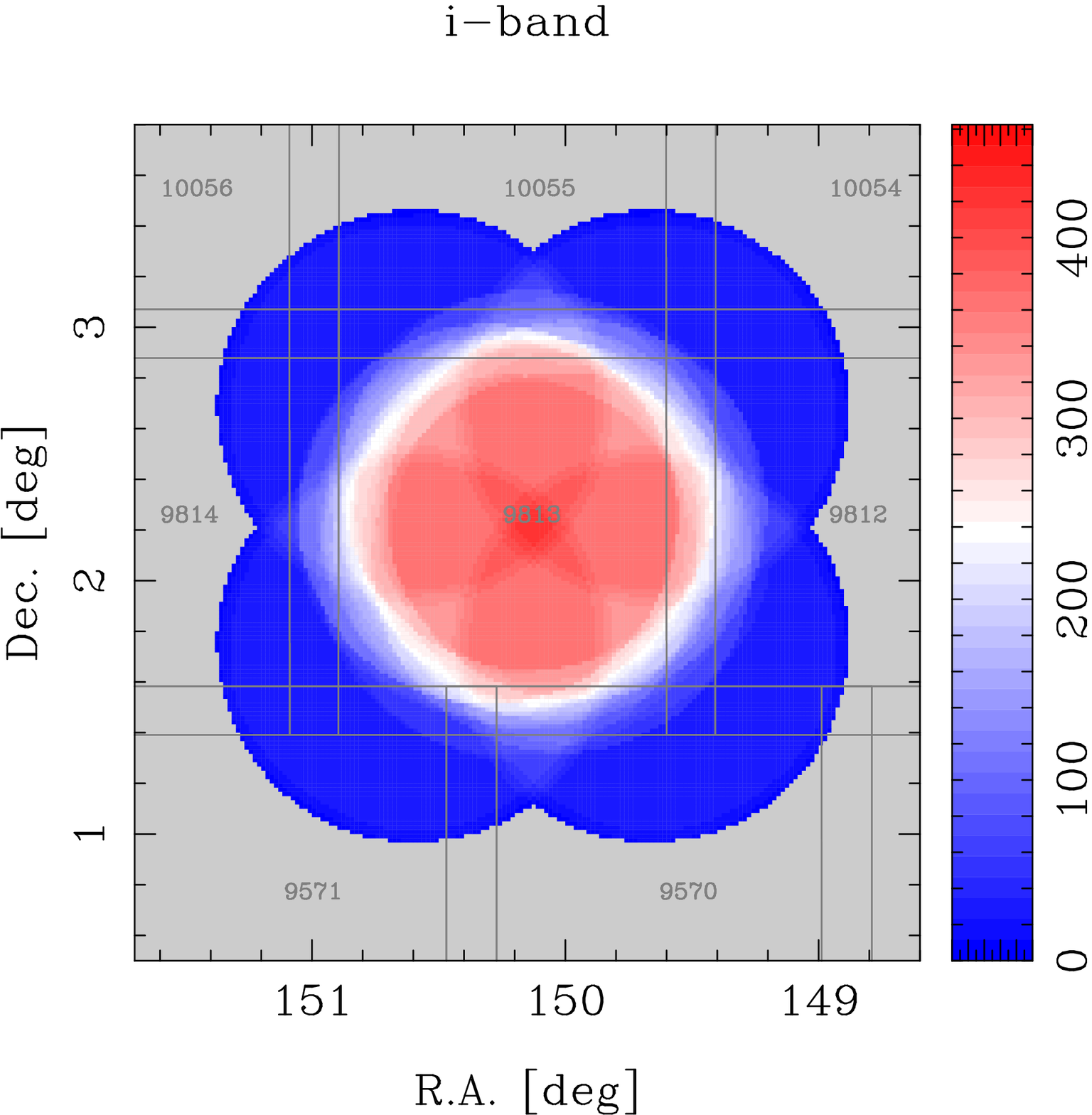}\hspace{0.5cm}
    \includegraphics[width=7cm]{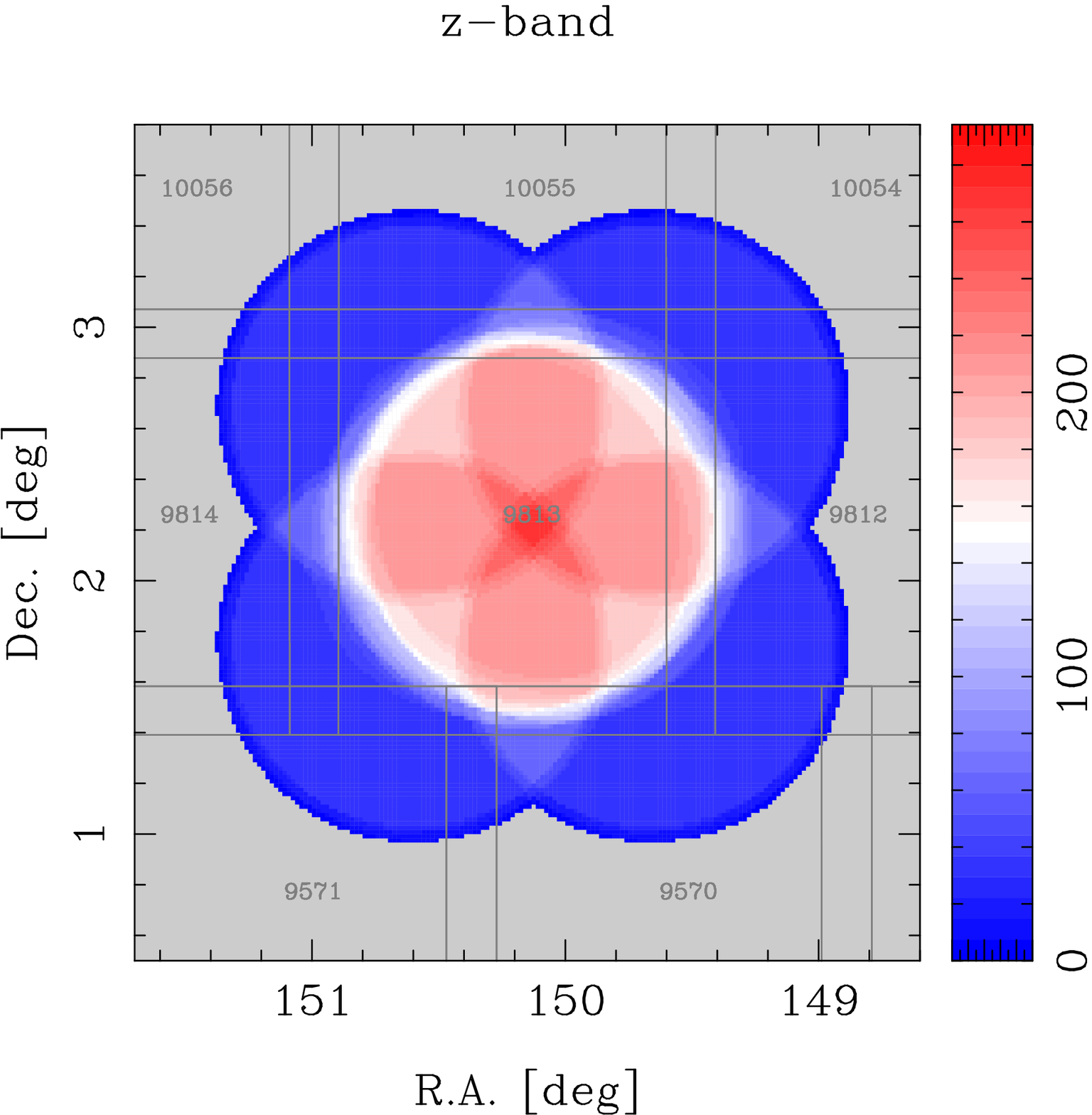}\\\vspace{0.5cm}
    \includegraphics[width=7cm]{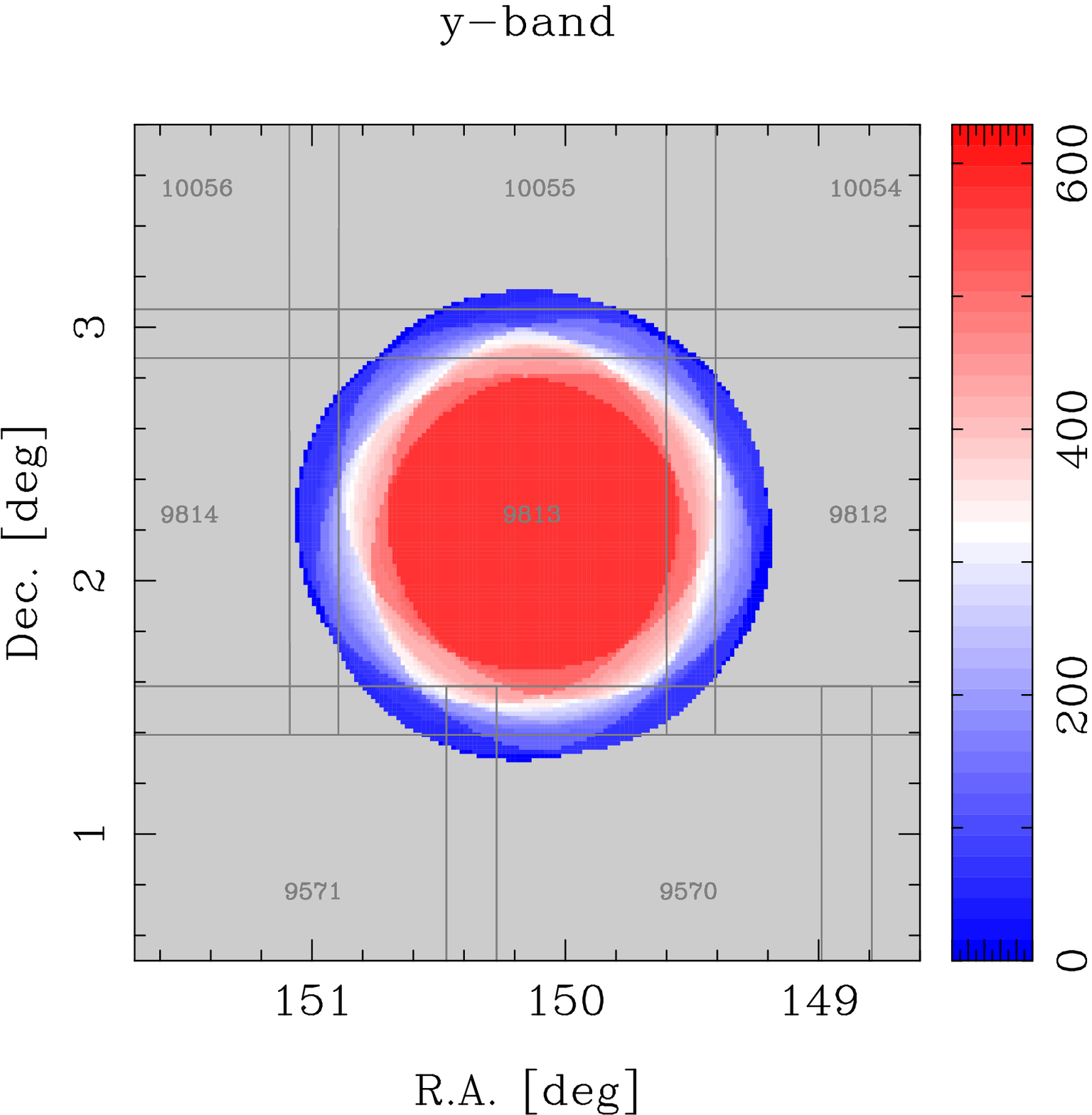}
  \end{center}
  \caption{
    Exposure map for each filter.  The color scale shows the exposure in minutes.
    Note that scale is different for different filters.
    The tract IDs and tract borders are shown in gray.
  }
  \label{fig:expmap}
\end{figure*}

We show in Fig.~\ref{fig:expmap} approximate exposure maps for each filter.
For simplicity, we approximate the HSC field of view as circular width of 1.5 deg diameter
and we ignore subtleties such as CCD gaps.  We apply only a small dither
in SSP and the resulting footprint is close to circular as shown in the $r$-band
plot, which just has the SSP data.
In the $giz$ bands, the UH data cover a wider region than the SSP data.  This introduces
inhomogeneities in the exposure map, but the data inside the COSMOS field are fairly uniform.
On the other hand, the $y$-band from UH has a more similar coverage to SSP.
As the $r$-band footprint is the smallest among the filters, this filter defines
the full-color area in the joint dataset.

\section{Quality Assurance}
\label{sec:qa}

We perform several tests in order to evaluate the quality of the data mostly  following
the methodology in \citet{aihara17a}.  We first start with seeing and depth
map for each filter as an overview of the dataset.  We then move on to perform
external astrometric and photometric tests using the PS1 catalog.
We also check internal photometric consistency using fluxes of bright stars measured
in different ways, and photometric zero-points using the location of the stellar sequence
on color-color diagrams.

\subsection{Seeing and depth maps}

\begin{figure*}
 \begin{center}
  \includegraphics[width=8cm]{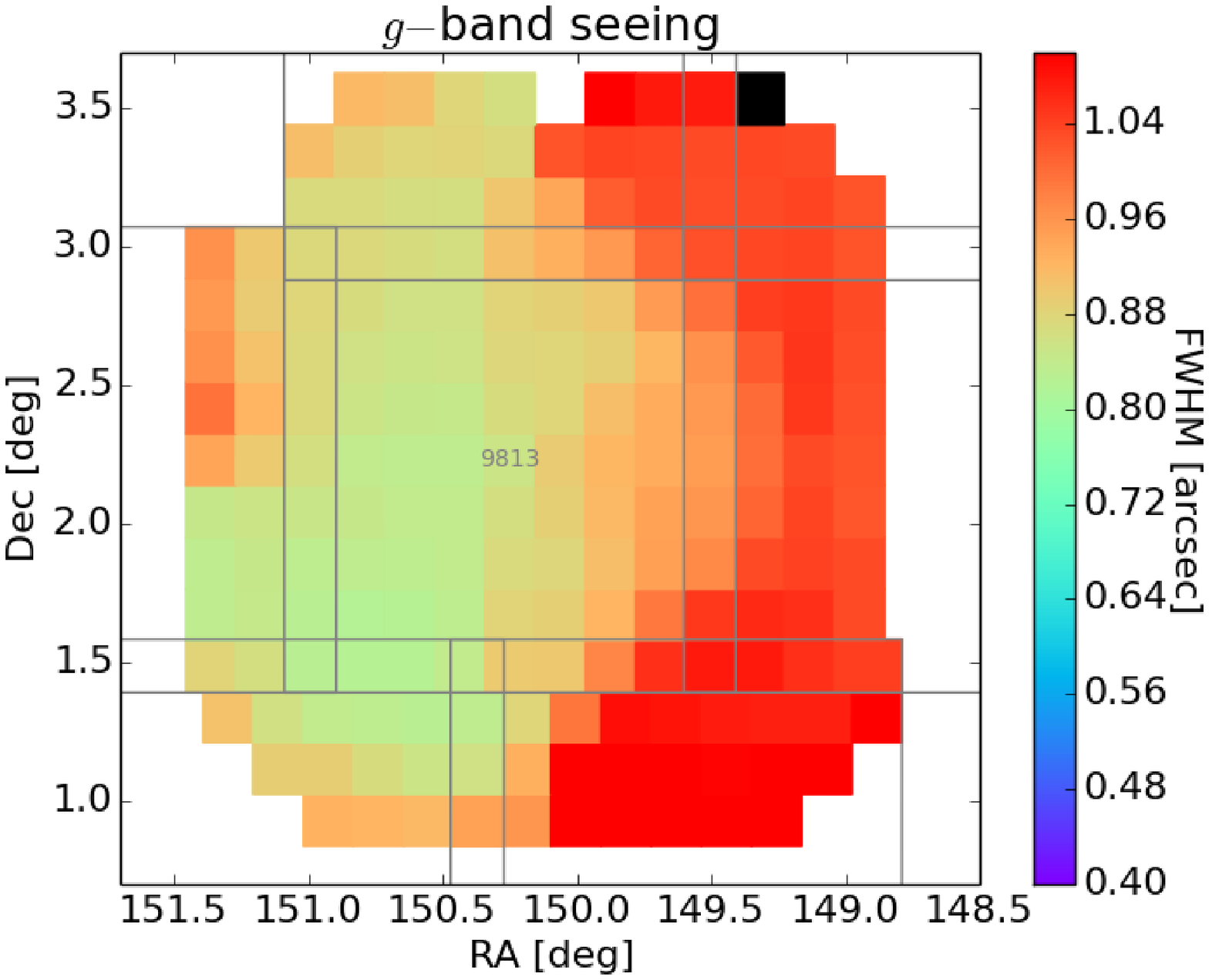}\hspace{0.5cm}
  \includegraphics[width=8cm]{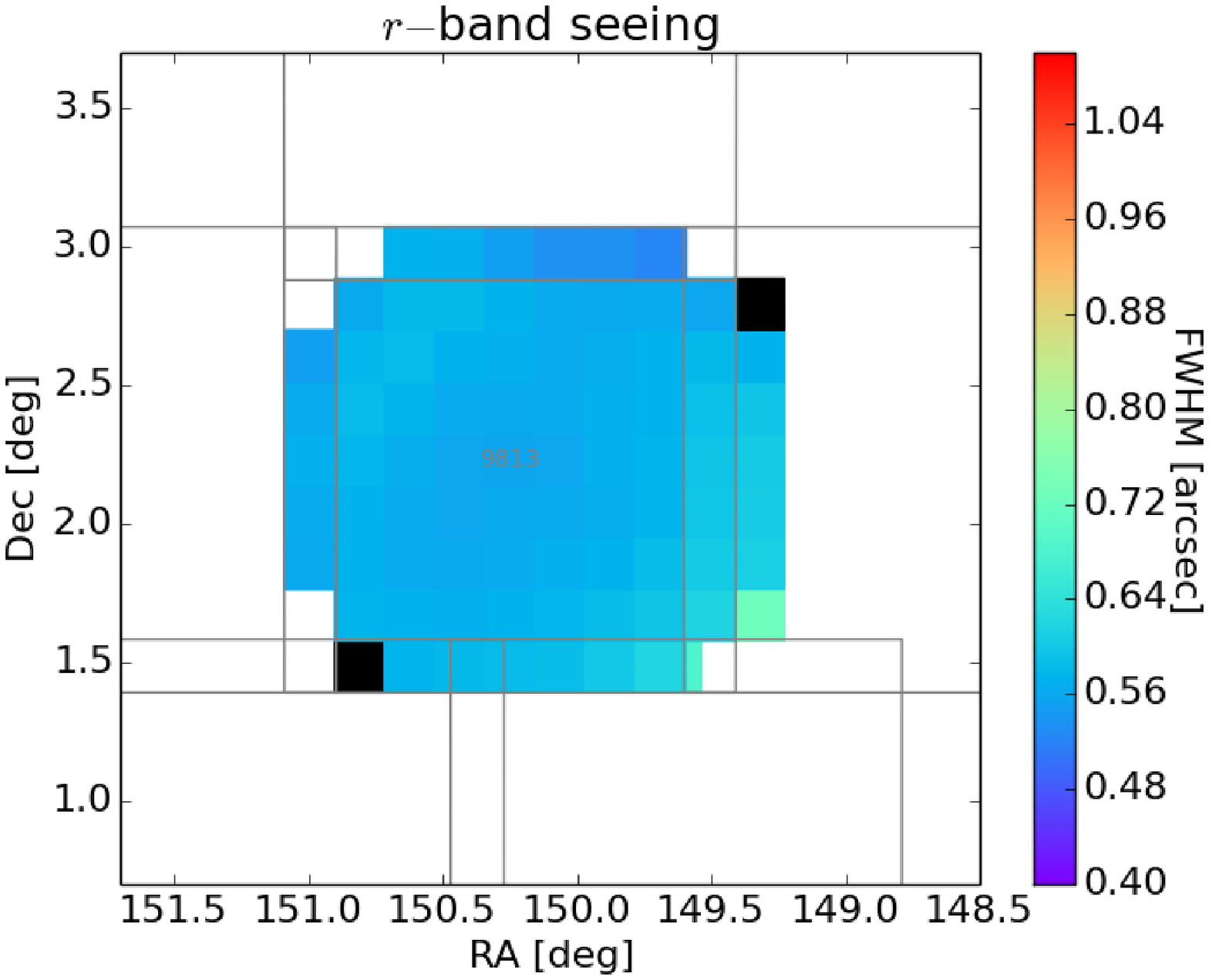}\\\vspace{0.5cm}
  \includegraphics[width=8cm]{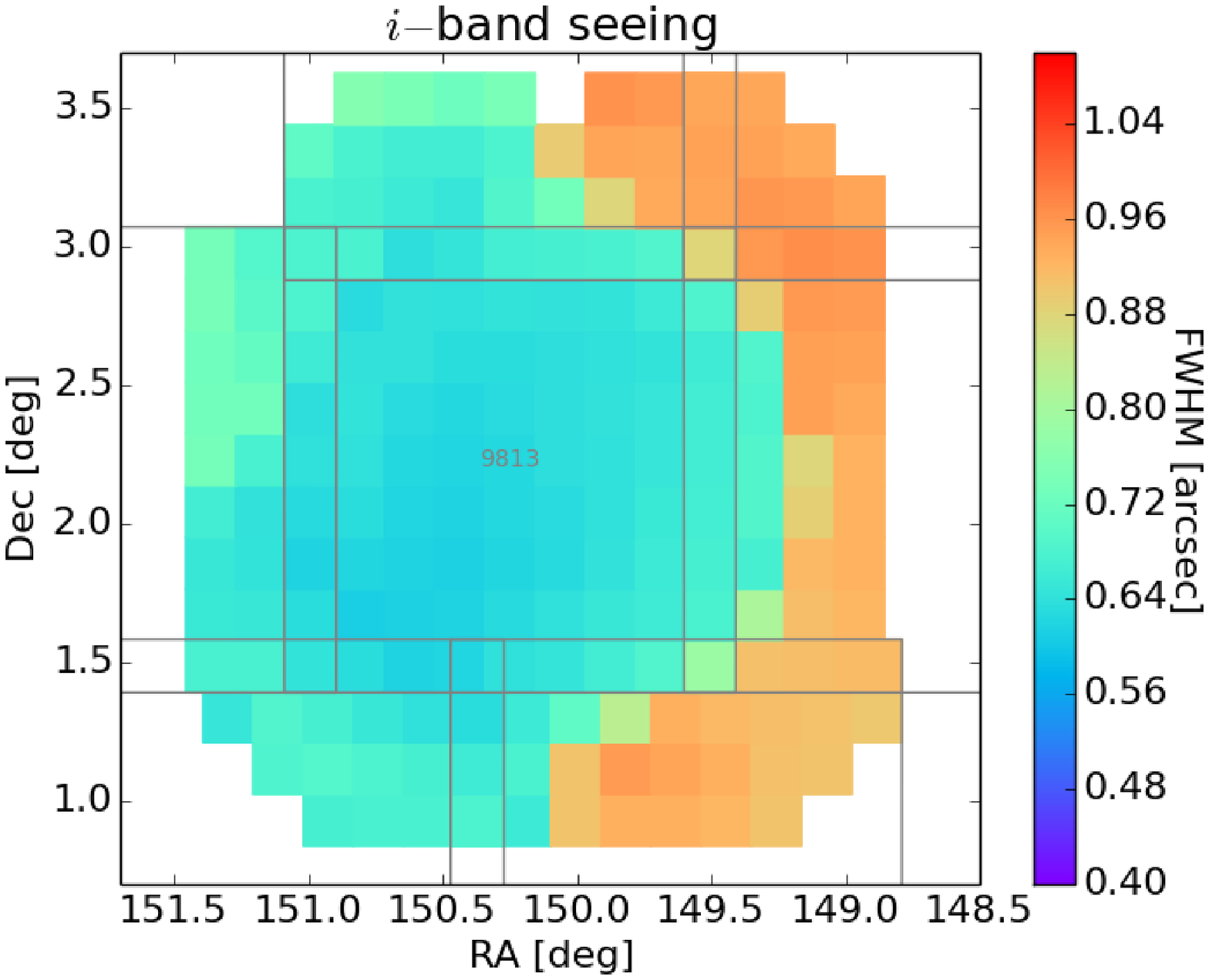}\hspace{0.5cm}
  \includegraphics[width=8cm]{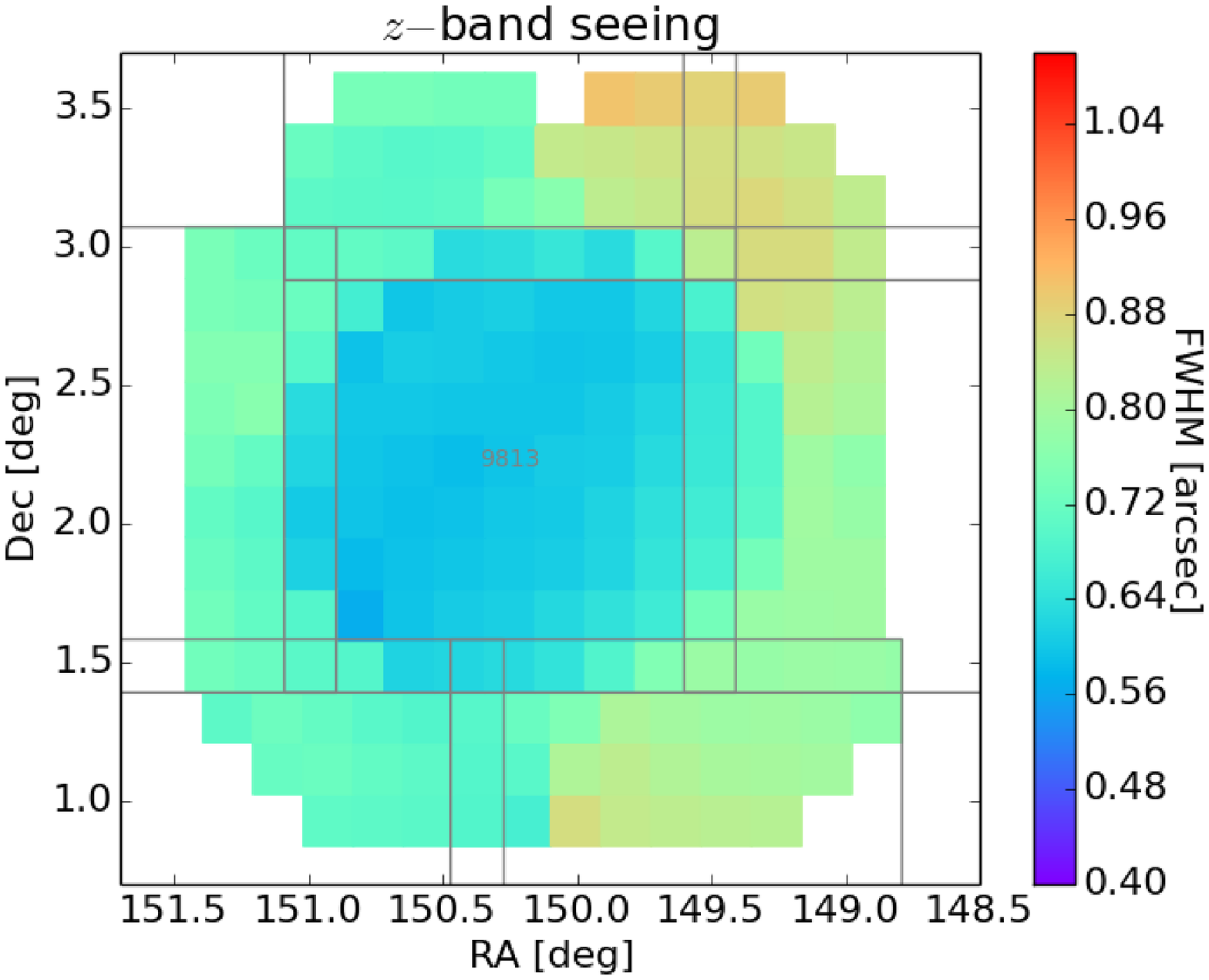}\\\vspace{0.5cm}
  \includegraphics[width=8cm]{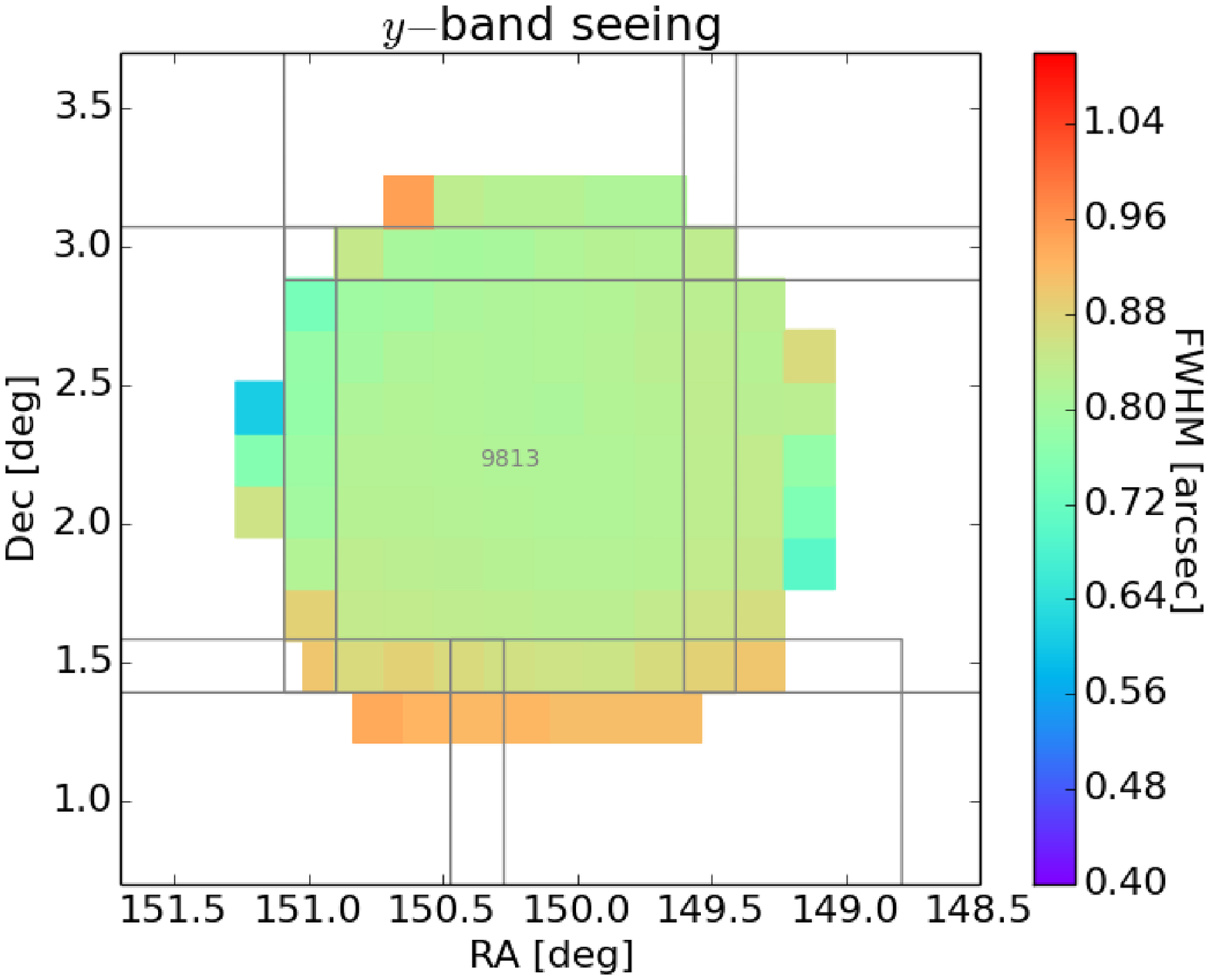}
 \end{center}
 \caption{
   Seeing for each filter.  The panels are for $grizy$ from top to bottom.
   The color bar on the right shows the FWHM scale.  Each
   square represents a patch and the gray lines show the tract borders.
 }
 \label{fig:seeing}
\end{figure*}

\begin{figure*}
 \begin{center}
  \includegraphics[width=8cm]{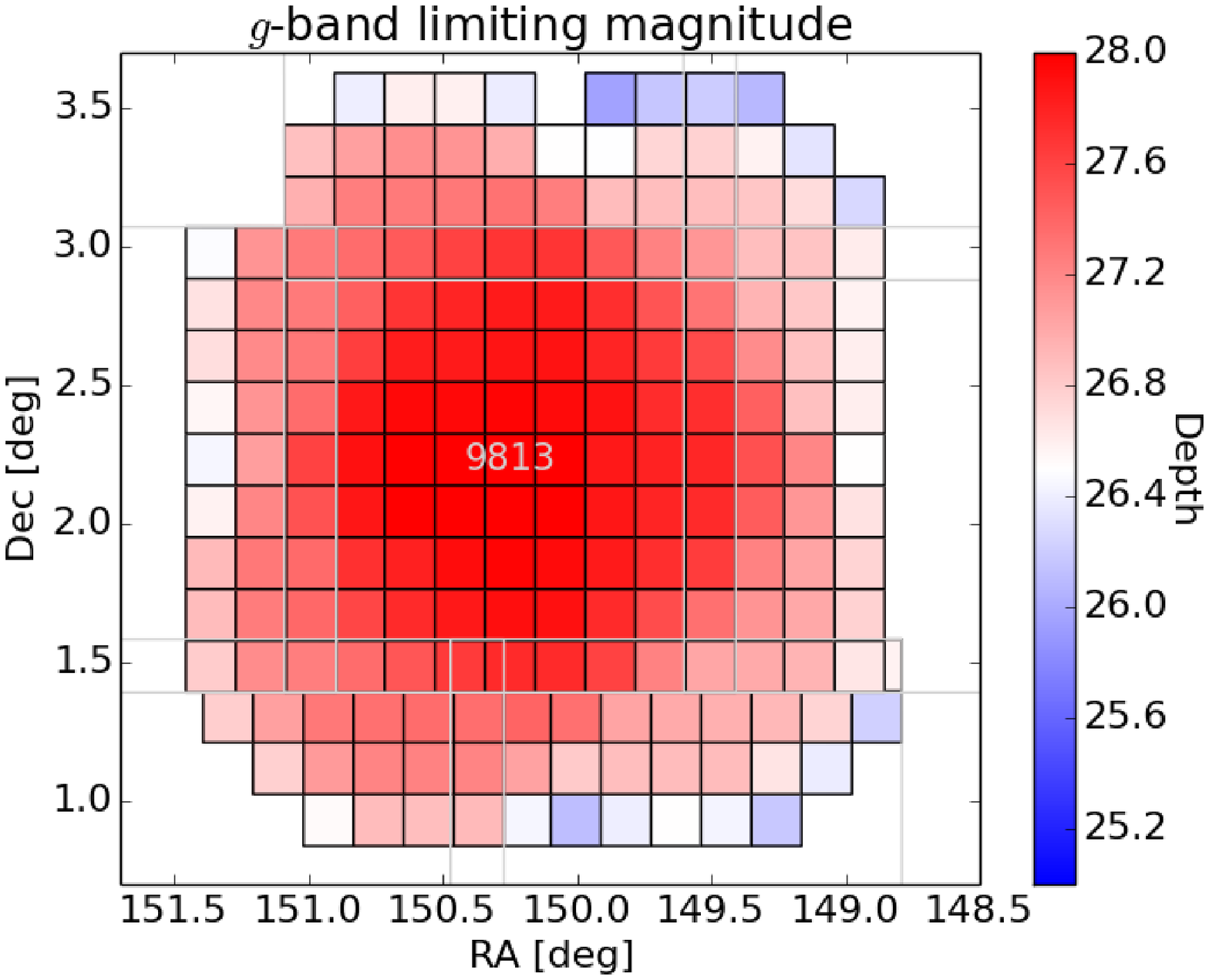}\hspace{0.5cm}
  \includegraphics[width=8cm]{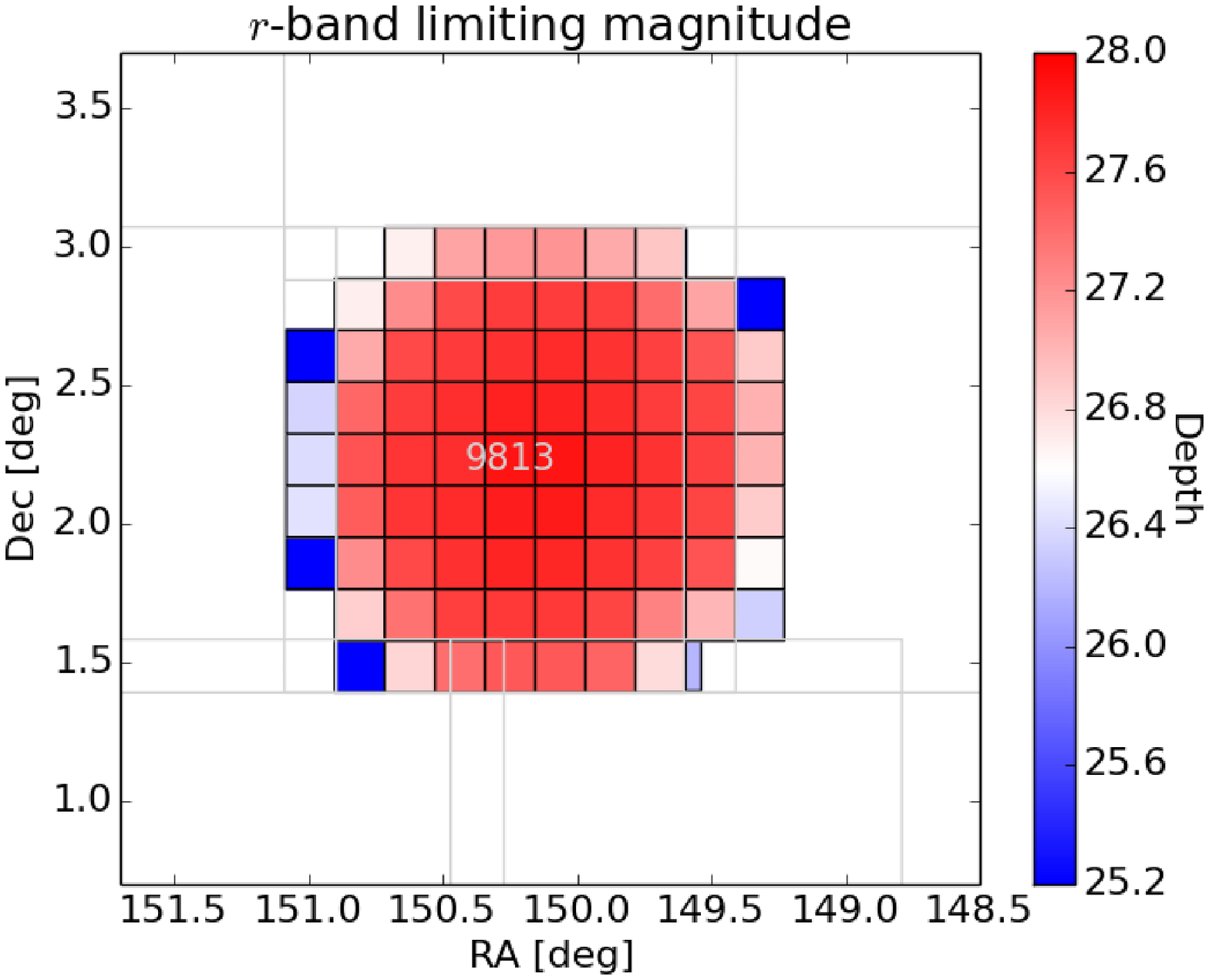}\\\vspace{0.5cm}
  \includegraphics[width=8cm]{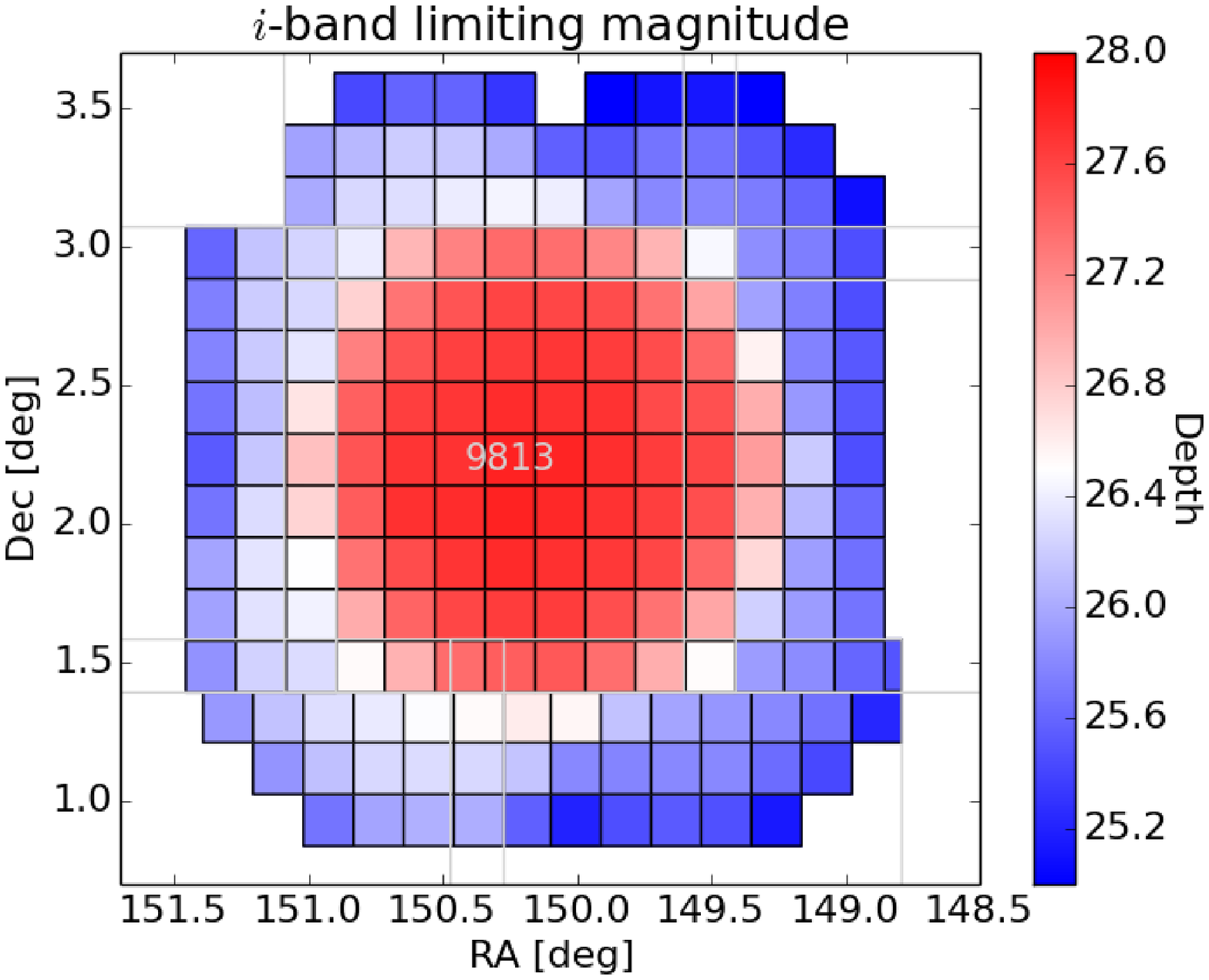}\hspace{0.5cm}
  \includegraphics[width=8cm]{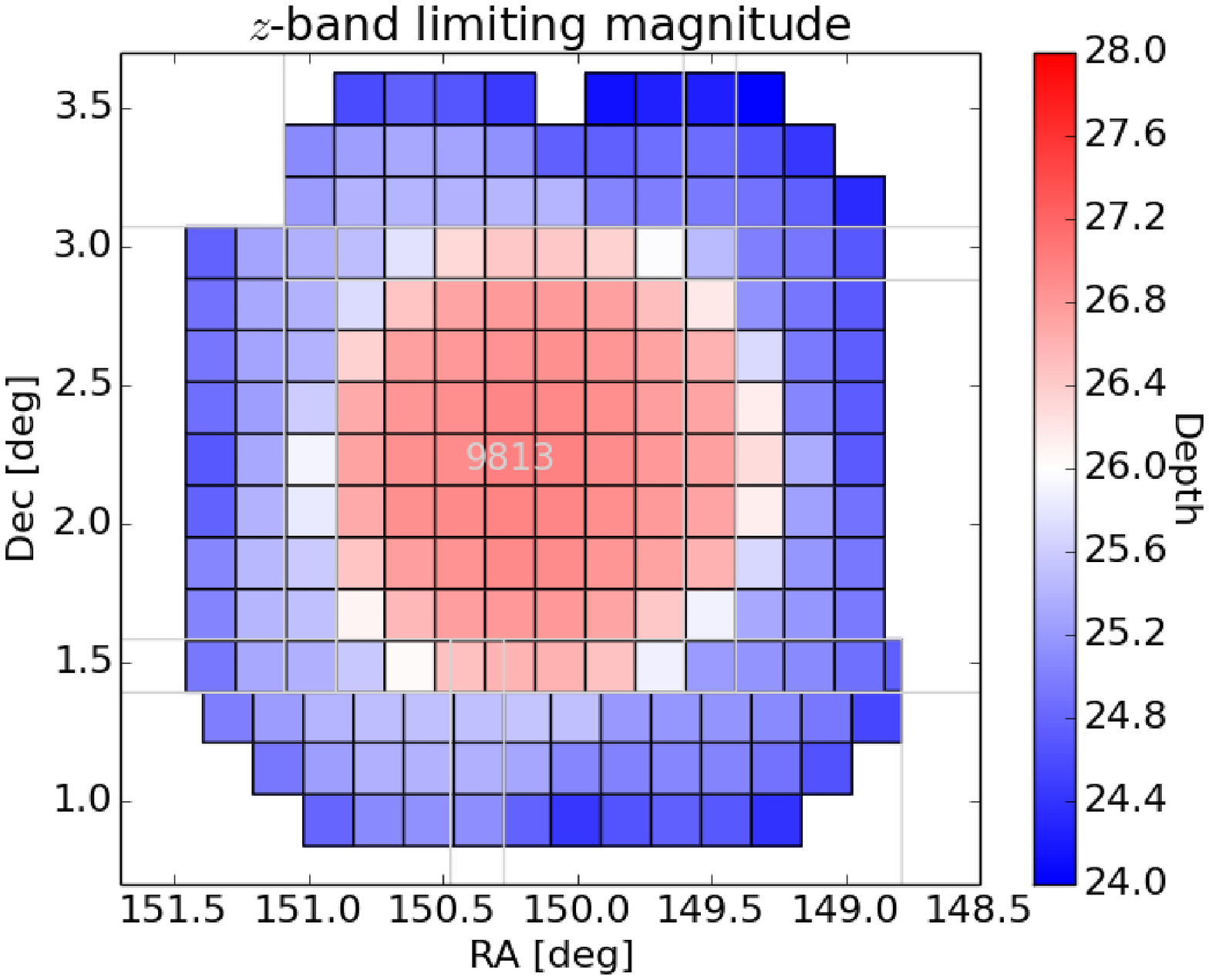}\\\vspace{0.5cm}
  \includegraphics[width=8cm]{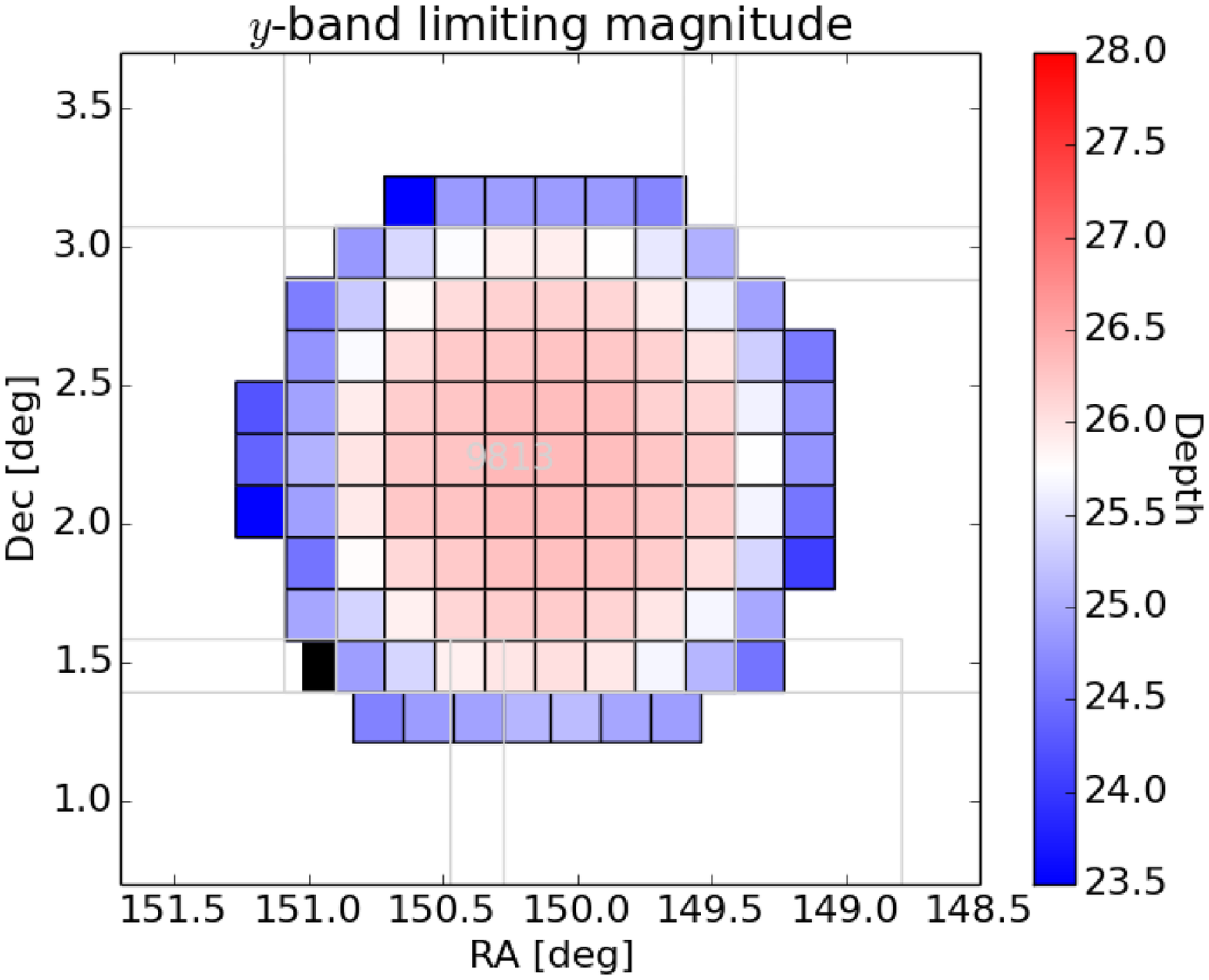}
 \end{center}
 \caption{
   Depth (5$\sigma$ point-source) for each filter.  The depths are based on the quoted
   photometric uncertainties by the pipeline and thus may be optimistic due to covariances.
   Each square represents a patch and the gray lines show the tract borders.
 }
 \label{fig:depth}
\end{figure*}

As mentioned earlier, the pointing coordinates and dither patterns are
different between HSC-SSP and UH, but the COSMOS field itself is covered well
by both datasets.  The  images are deepest in the central tract 9813,
which covers most of COSMOS.
Typical integration times, seeing, and depths in this tract
for each filter are summarized in Table~\ref{tab:data}.
The exposure times range between 1.5 and 9.5 hours and the red filters
have longer integration times than the blue ones.  The seeing is 0.6 to 0.9 arcsec.
Note that the seeing is derived from Gaussian-weighted second moments \citep{aihara17a,bosch17}.
Fig.~\ref{fig:seeing} shows the spatial distribution of the seeing in each filter.
The $g$- and $i$-bands have some structure, while the other filters are fairly
homogeneous.  Thanks to the good seeing, we achieve depths of
26.2~mag in the $y$-band to 27.8~mag in the $g$-band\footnote{
  These depths are based on PSF flux uncertainties quoted by the pipeline.
  Due to the covariances between the pixels, the estimated depths are likely overly
  optimistic by up to $\sim0.5$~mag.
}.
Compared to DR1,
the joint dataset is deeper by up to 0.4~mag.  These are the deepest optical
images of the COSMOS field available to date. 
Fig.~\ref{fig:depth} shows the spatial variation of the depths.
The depth maps show similar structure to the seeing
map because the depths are for point sources, but the spatial variations of
the depths are fairly small within the COSMOS field, which is mostly covered
in tract 9813.

\begin{table*}[htbp]
  \begin{center}
    \begin{tabular}{cccc}
      \hline\hline
      filter & integration time (min.) & seeing FWHM (arcsec) & depth\\
      \hline
      $g$    & 140 (85)                & 0.92 (0.82)          & 27.8 (27.5)\\
      $r$    & 90  (90)                & 0.57 (0.57)          & 27.7 (27.7)\\
      $i$    & 360 (160)               & 0.63 (0.63)          & 27.6 (27.2)\\
      $z$    & 210 (147)               & 0.64 (0.62)          & 26.8 (26.7)\\
      $y$    & 570 (250)               & 0.81 (0.74)          & 26.2 (25.8)\\
      \hline\hline
    \end{tabular}
  \end{center}
  \caption{
    Typical integration times, seeing, and $5\sigma$ depths for point sources
    in tract 9813.   The numbers in parentheses are for HSC PDR1 for comparison.
  }
  \label{tab:data}
\end{table*}

\subsection{Astrometric and Photometric quality}

We turn our attention to the overall astrometric and photometric accuracy.
We first compare astrometry against PS1.  As we have used the PS1 catalog to
calibrate our astrometry in individual CCDs in the pipeline processing \citep{aihara17a,bosch17},
this is not strictly an external comparison, but it is still useful to check
the astrometry on the coadds.

\begin{figure*}
 \begin{center}
   \includegraphics[width=8cm]{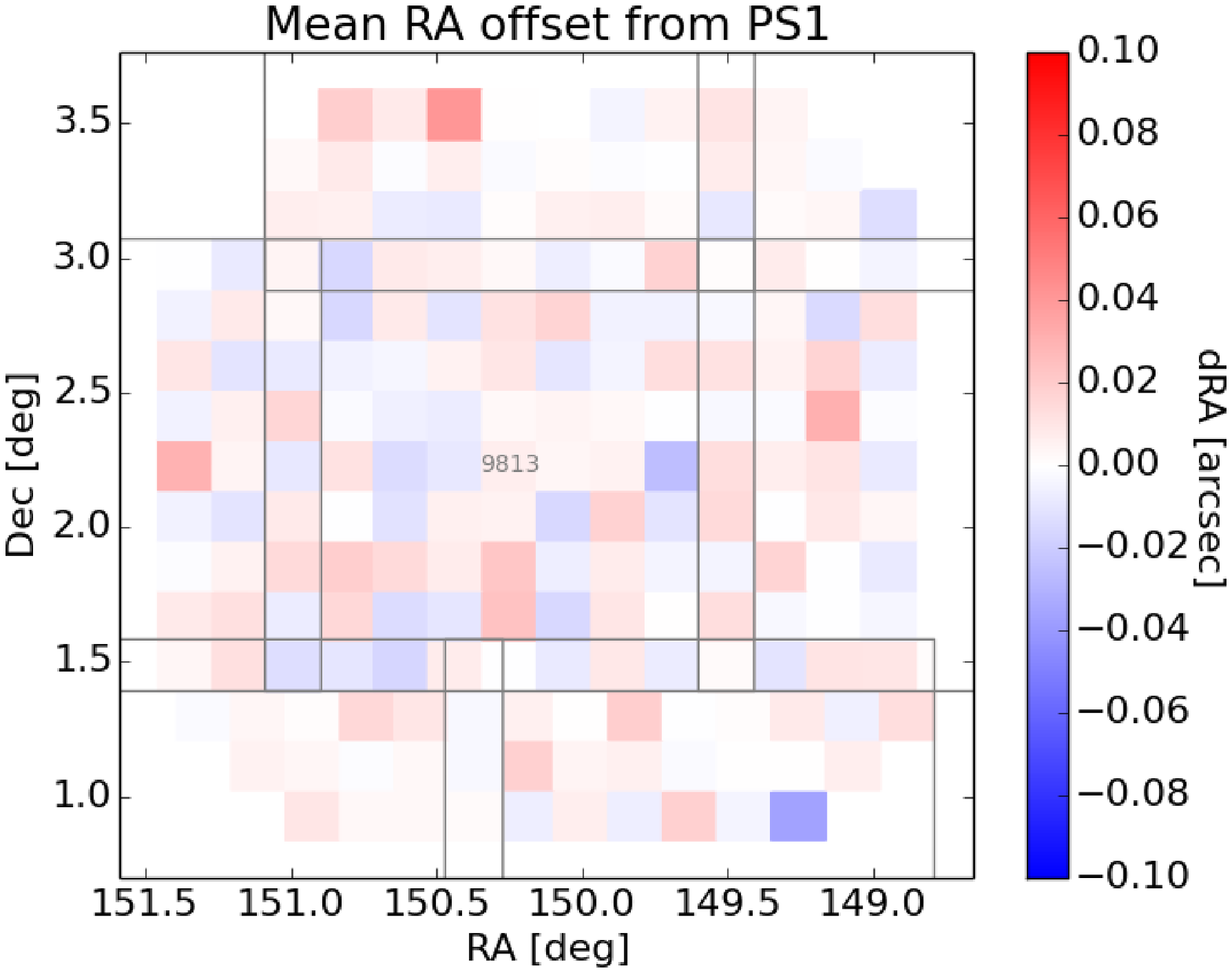}\hspace{0.5cm}
   \includegraphics[width=8cm]{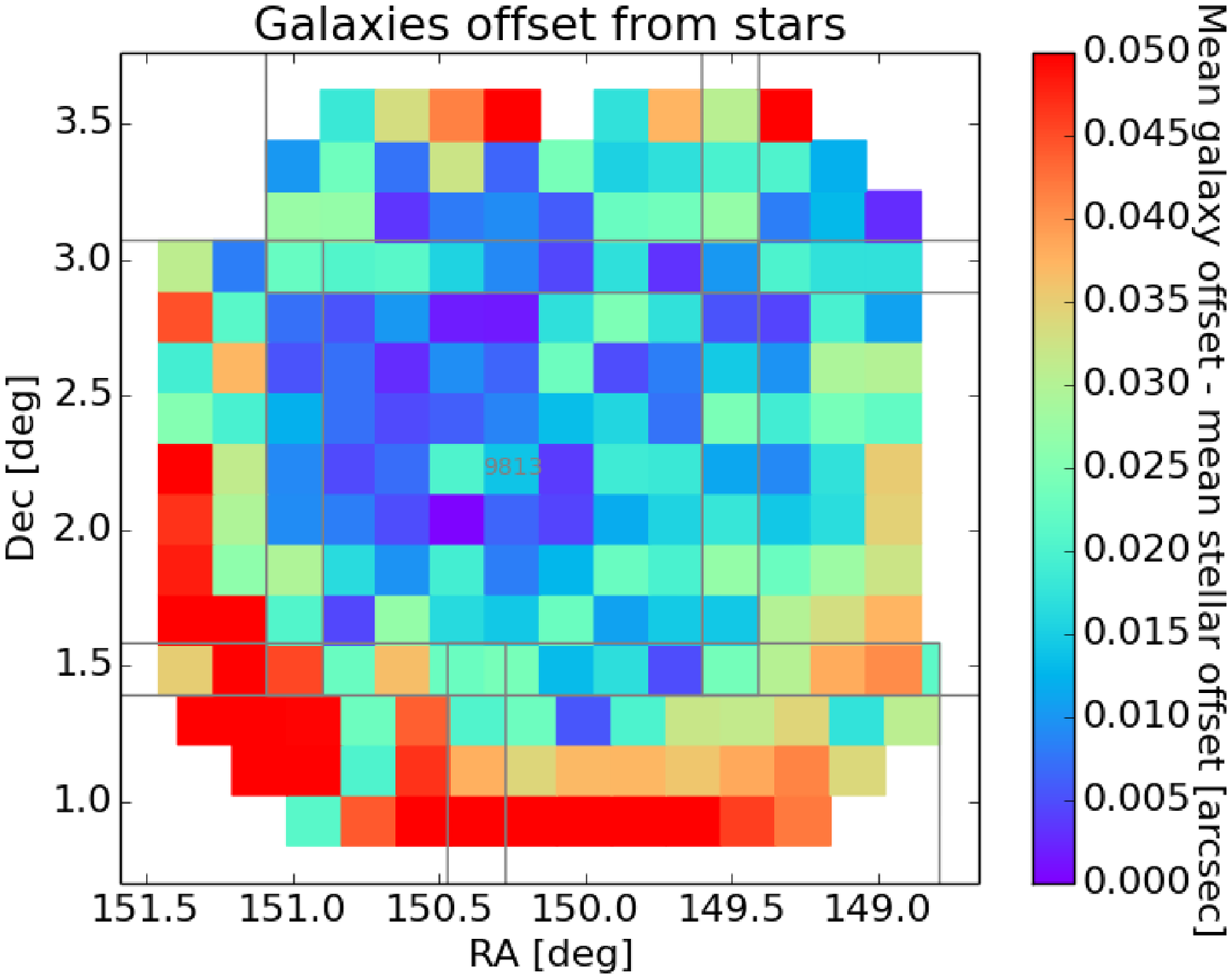}
 \end{center}
 \caption{
   The left panel shows the mean R.A. offset per patch against the PS1 reference
   catalog in the $i$-band.  The scale on the right shows the offset in arcsec.
   The right panel shows the difference in the mean offset between galaxies and stars.
   Each rectangle represents a patch and the lines show the tract borders.
 }
 \label{fig:astrometry}
\end{figure*}

The left panel of Fig.~\ref{fig:astrometry} shows the mean R.A. offset of stars brighter than
20th mag in the $i$-band between HSC and PS1.  This is one of the astrometric
quality measures and more plots are available at the data release site.
The astrometry is good to 30~mas across the field and there is no clear spatial structure
to the residuals.
The offset in Dec. (not shown here) looks very similar.
The right panel gives the mean positional offset between galaxies and stars.
The bottom-left corner of the field shows a somewhat large
($\sim0.05$ arcsec) offset.  This could be due to the fact that the PS1 PV2 catalog
 is referenced to 2MASS and the proper motions are ignored.  The recently
released PV3 catalog \citep{berghea16} may solve the problem.
The mean statistics over the entire observed field for all the filters are
summarized in Table~\ref{tab:astrometry}.

\begin{table}
  \begin{center}
    \begin{tabular}{cccc}
      \hline
      Filter & \centering RA vs PS1 & Dec vs PS1 & Star-Galaxy offset \\
      & \multicolumn{1}{c}{(mas)} & \multicolumn{1}{c}{(mas)} & \multicolumn{1}{c}{(mas)} \\
      \hline
      $g$ & $31$ & $32$ & $ 6$\\
      $r$ & $32$ & $29$ & $ 6$\\
      $i$ & $31$ & $30$ & $10$\\
      $z$ & $34$ & $33$ & $11$\\
      $y$ & $34$ & $32$ & $14$\\
      \hline
    \end{tabular}
  \end{center}
  \caption{
    Astrometric quality.
    The first two statistical columns are the RMS of residuals of the stated quantity against PS1 for stars
    brighter than 20th mag.   The last column is the mean of the residual offset against PS1 between stars and galaxies.
  }
  \label{tab:astrometry}
\end{table}

\begin{figure*}
  \begin{center}
    \includegraphics[width=8cm]{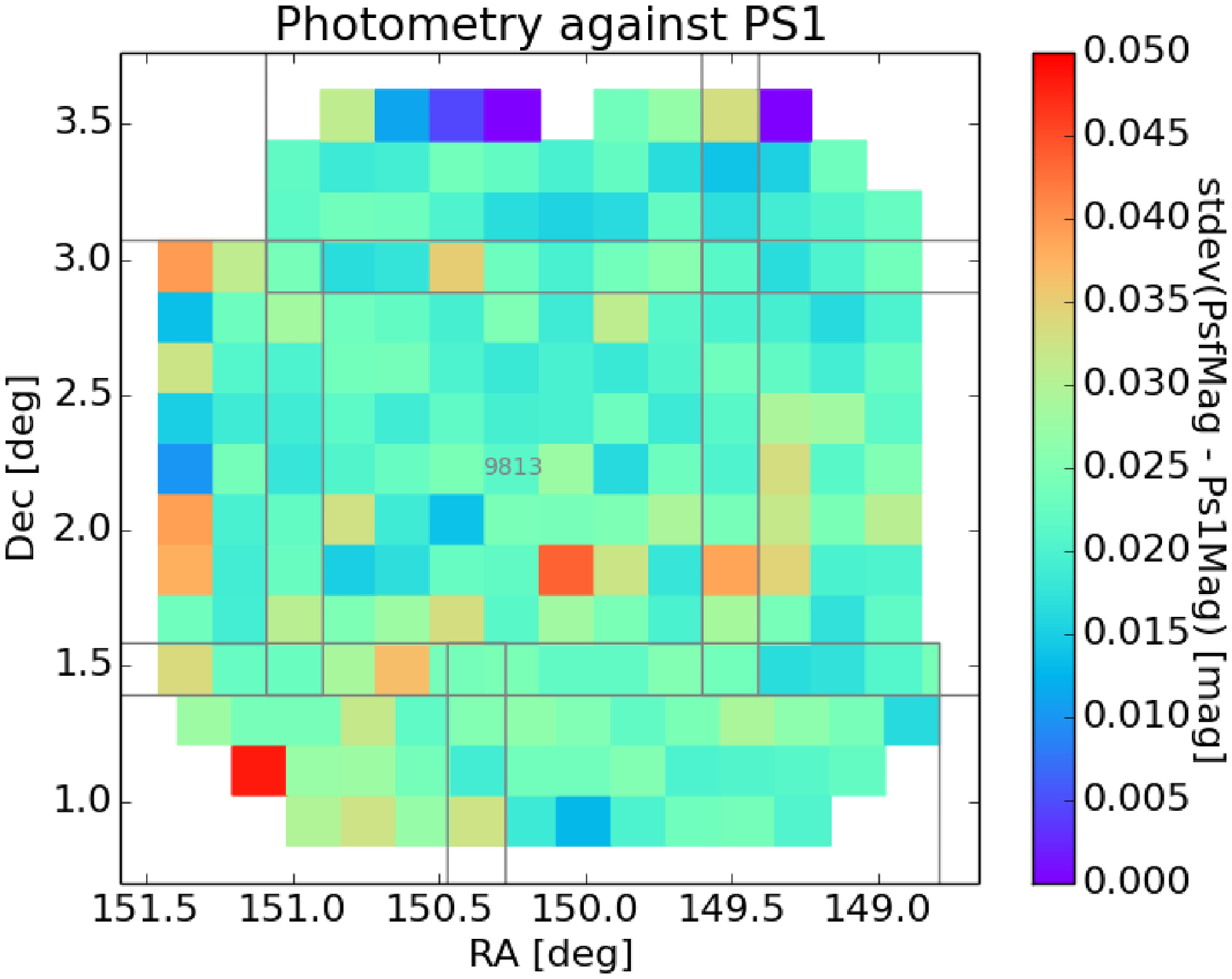}\hspace{0.5cm}
    \includegraphics[width=8cm]{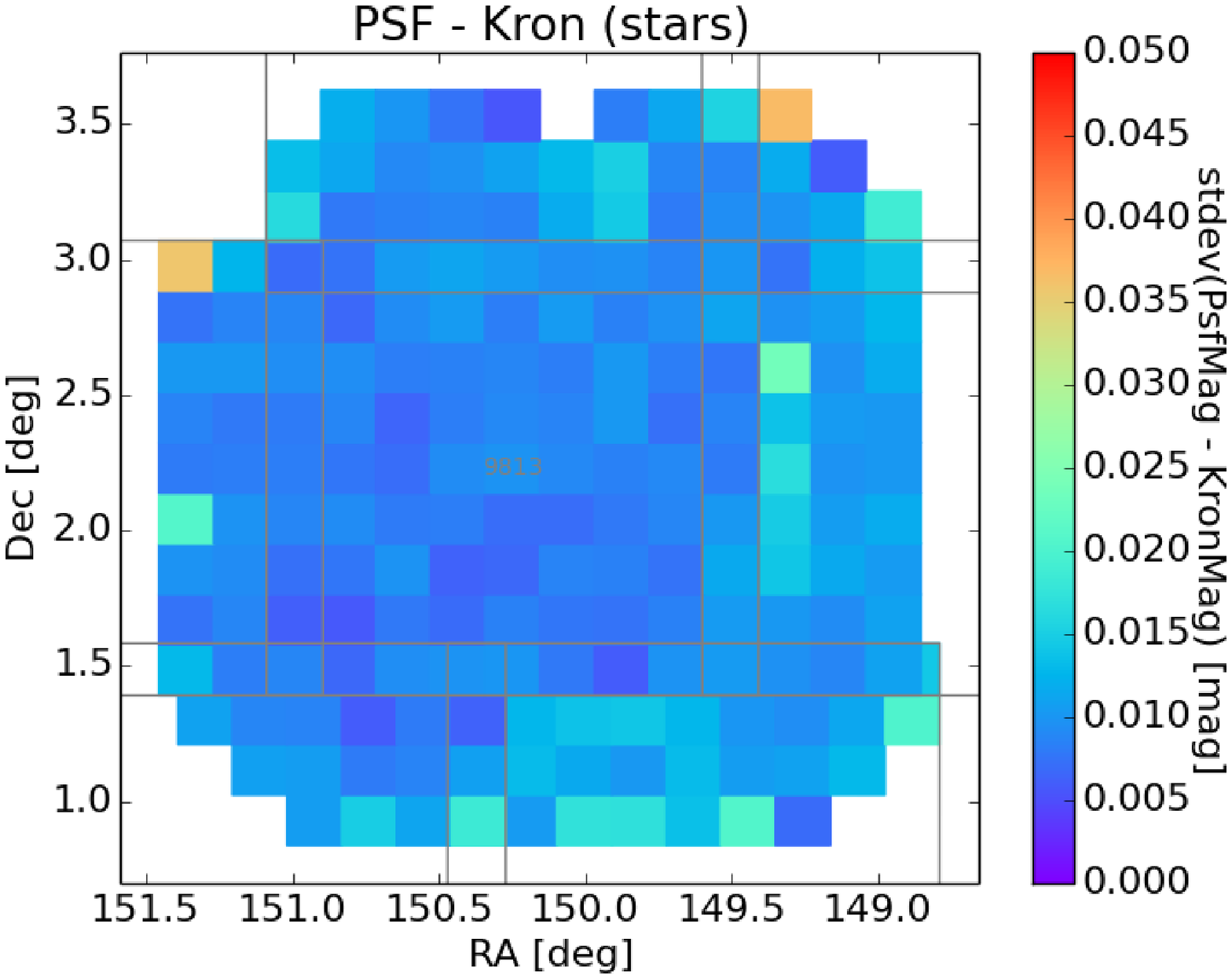}
  \end{center}
  \caption{
    The left plot shows the RMS of the difference in stellar PSF magnitudes between
    HSC and PS1 in the $i$-band.  Stars brighter than $i=20$ are used here.
    The right plot shows the RMS of the difference between the PSF and
    Kron magnitudes.  Both magnitudes are from HSC and stars brighter than $i=21.5$ are
    included in the statistics.  Each rectangle represents a patch and the gray lines show
    the tract borders.
  }
  \label{fig:photometry}
\end{figure*}

Next, we make both external and internal tests for photometry.  The left panel of
Fig.~\ref{fig:photometry} compares the $i$-band PSF photometry between HSC and PS1
for stars brighter than $i_{PSF}=20$.  The scatter is typically 0.02~mag, representing
the quadrature sum of uncertainties from both surveys.  This scatter is not superb, but
is good enough to enable many science goals.  For internal consistency, we compare
PSF mags and Kron mags for stars brighter than $i_{PSF}=21.5$.  Because we are
measuring the same objects with different techniques, we expect the difference
to be small.  We find the RMS scatter is about 0.01~mag, which indeed suggests
good internal consistency. More plots can be found online,
and they all indicate reasonably good photometric accuracy.
Table~\ref{tab:photometry} summarizes the numbers for all the filters.

\begin{table}
  \begin{center}
    \begin{tabular}{cccc}
      \hline
      Filter & \centering PSF vs PS1 & PSF - Kron & PSF - CModel \\
      & \multicolumn{1}{c}{(mmag)} & \multicolumn{1}{c}{(mmag)} & \multicolumn{1}{c}{(mmag)} \\
      \hline
      $g$ & $19.1$ & $11.3$ & $3.2$\\
      $r$ & $22.4$ & $10.8$ & $2.9$\\
      $i$ & $23.9$ & $11.0$ & $2.6$\\
      $z$ & $18.0$ & $15.0$ & $2.3$\\
      $y$ & $32.3$ & $13.0$ & $1.5$\\
      \hline
    \end{tabular}
  \end{center}
  \caption{
    Photometric quality.
    The first statistical column is the RMS of residuals of the PSF magnitudes for stars
    brighter than 20th mag.  The last two columns are the RMS of the difference between the two stated magnitudes for
    stars brighter than 21.5~mag.
  }
  \label{tab:photometry}
\end{table}

\begin{figure*}
 \begin{center}
  \includegraphics[width=8cm]{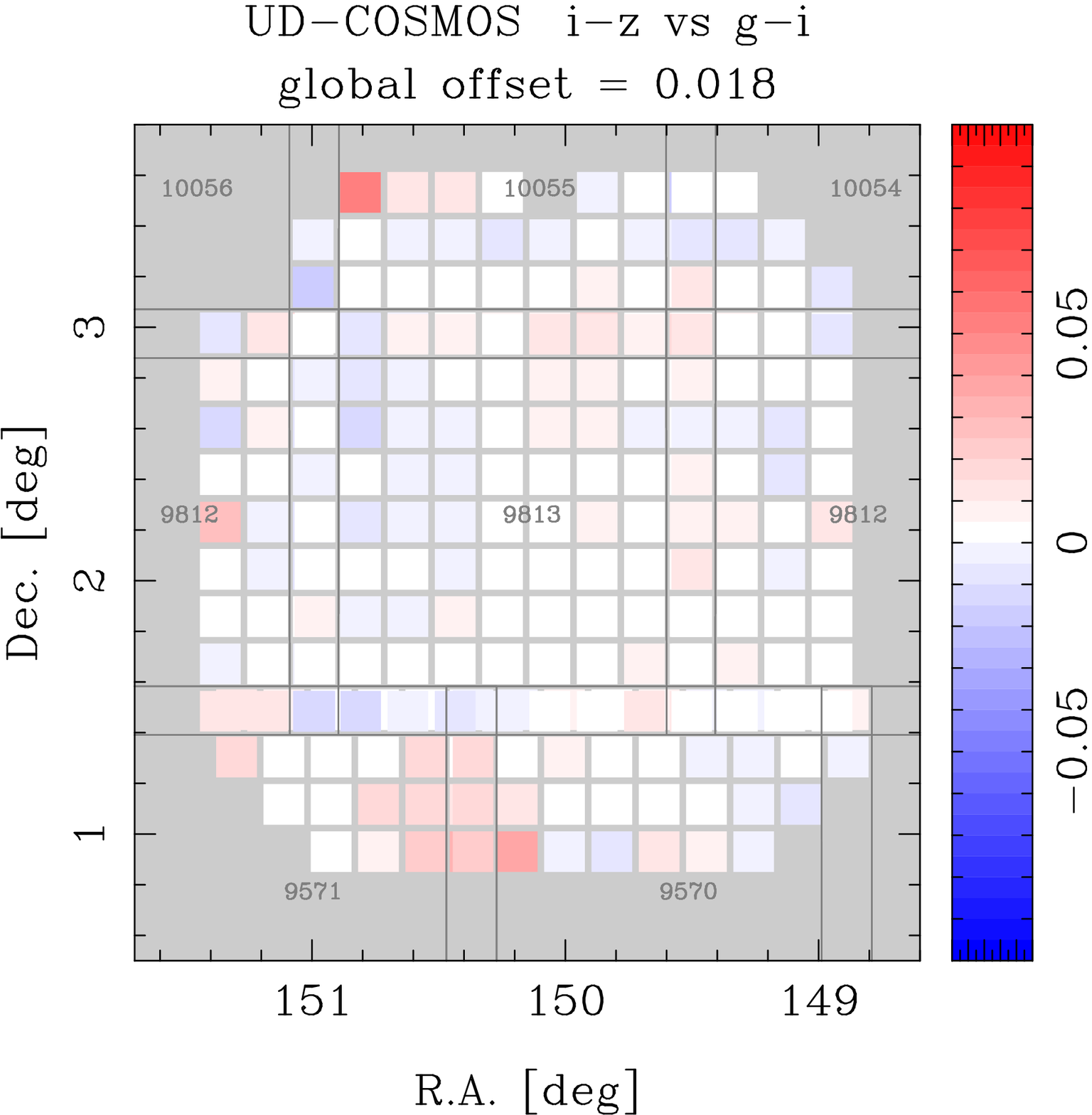} 
  \includegraphics[width=8cm]{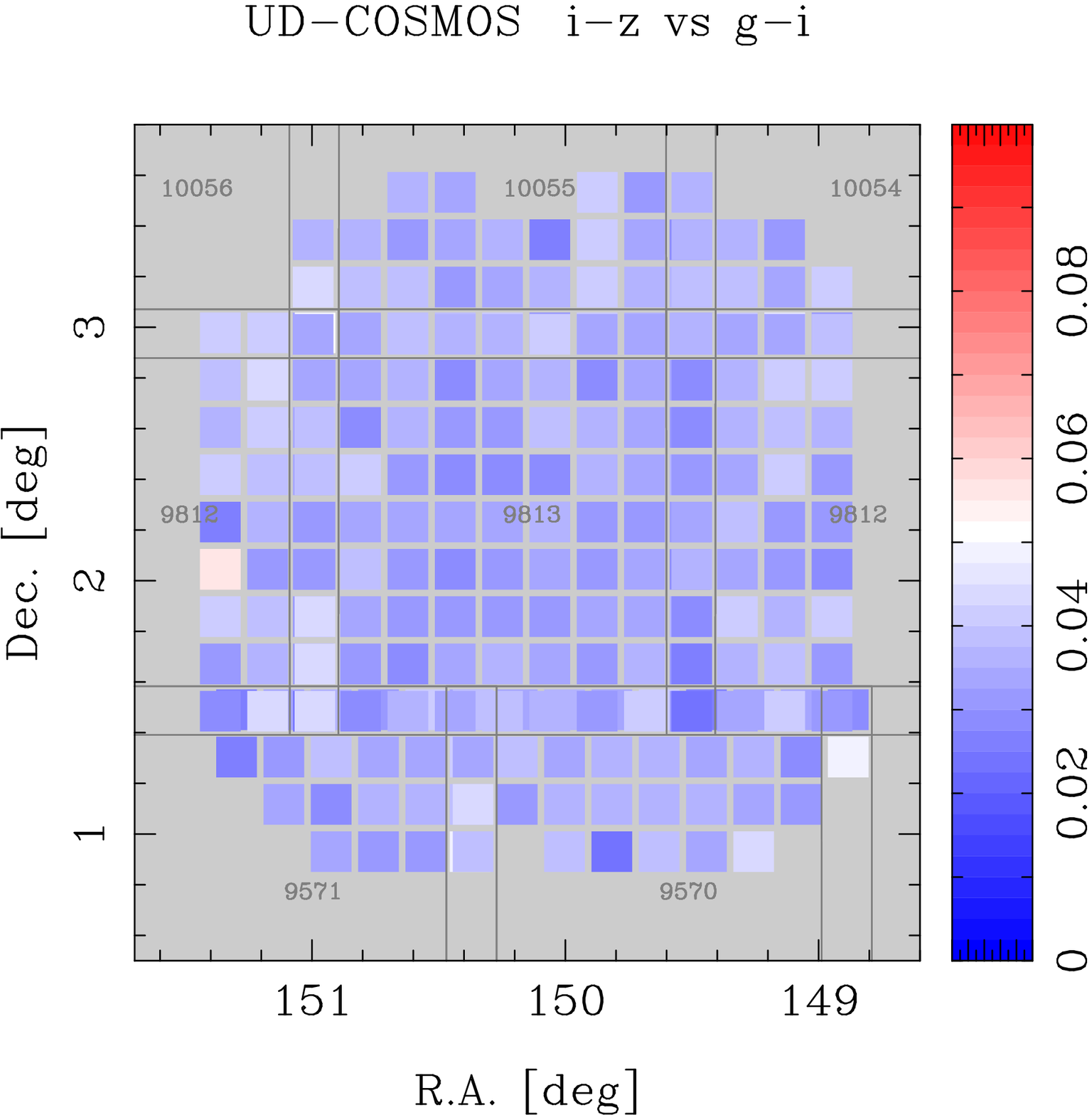} 
 \end{center}
 \caption{
   Color offset in the stellar sequence (left) and color scatter (right) on
   the $i-z$ vs. $g-i$ diagram.
   The median color offset across the field (0.018~mag) is subtracted
   to highlight the spatial inhomogeneities.
   The tract IDs and tract borders are shown in gray.
 }
 \label{fig:stellar_sequence}
\end{figure*}

Finally, we test the spatial uniformity of the photometric zero-points across
the field.  We estimate an offset between the location of the observed stellar
sequence and that of the synthetic \citet{gunn83} stellar sequence on
a color-color diagram.  As in the previous tests, we use only bright stars
($i_{PSF}<21.5$) to ensure that statistical uncertainties
do not dominate the error budget.
An offset in the stellar sequence is degenerate between the two colors chosen,
and we make an assumption that the offset is entirely in the y-direction.
We apply corrections for the Galactic extinction, but not all the stars
are behind the Milky Way's dust screen, which may introduce an additional
offset and scatter.  Fig.~\ref{fig:stellar_sequence} shows the spatial
variation of the zero-point for one combination of colors.  Note that,
in order to enhance the spatial inhomogeneities, we remove the global offset,
which can be due to inaccuracies in our system response functions, the \citet{gunn83}
spectra, Galactic extinction correction, etc.
The figure suggests that the zero-point is fairly uniform across the field
at a 1\% level.  The scatter shown in the right panel is about 3\%, but this
is due to three filters and if we divide by $\sqrt{3}$, it is about 2\%
in each filter.

To summarize, our photometry is accurate to about 1-2\% over the entire
COSMOS field.  This should be good enough for many extragalactic science cases,
especially for exploration of faint galaxies in
the distant Universe in this unique extragalactic field.

\section{Summary and Data Access}
\label{sec:summary}

We have presented the joint COSMOS dataset taken by the SSP and UH teams.
We reach $\sim27.5$ mag at $5\sigma$ for point sources with good image quality of
0.6-0.9 arcsec.  The astrometry is accurate to 30~mas and photometry to $\sim2$\%.
These are the deepest COSMOS images available in the optical
wavelengths and will be extremely useful to explore, e.g., the high-$z$ Universe
with unprecedented statistics.  The COSMOS field is one of the UltraDeep fields
of the SSP survey and the SSP team is collecting more data there.  We will
eventually reach $\sim1$~mag deeper.

The joint dataset is served at the SSP data release site (\url{https://hsc-release.mtk.nao.ac.jp}) and both the catalog and
image products can be easily retrieved via database or direct download.
Our database offers the easiest way to retrieve the catalog data.  Users are
referred to \citet{aihara17a} and online manual such as schema browser
about the database tables.  As described in \citet{bosch17}, each measurement
algorithm comes with a flag to indicate measurement success/failure.  Also,
objects in the overlapping regions between the adjacent patches and
tracts are duplicated in the database as they are multiply detected and measured
by the pipeline.  Users should apply flag cuts in order to select objects
with clean photometry and to remove the duplicates.  A suggested set of
flags is given in \citet{aihara17a}.

Both coadd and individual CCD images are available for direct download.
There is an online search tool, which allows users to specify a sky region,
filter, and data products to download.  There is also an image cutout tool,
which accepts user upload to generate postage stamps of many objects.
The online image browser, hscMap, will be the most useful tool for quick
browsing of the images.  See the online manual for all these tools.

All of the image products from the pipeline are in the fits format and have science,
mask, and variance images in separate HDUs.  The meanings of the mask bits
are explained in the header.  The coadd images are likely the most interesting
data products to the community and they are available for each patch.  In order
to generate a tract-wide image, we offer a script to combine the patch images.
The coadds have a homogenized photometric zero-point of 27.0~mag/ADU, but
aperture corrections need to be applied in order to obtain accurate fluxes of objects.
See \citet{bosch17} for details.

We would like to ask users to acknowledge the SSP and UH teams when using this
joint dataset in publications.
We also encourage users to reference relevant technical papers from HSC-SSP
as well as this paper.  The suggested acknowledgment text and a list of
the technical papers can be found at the data release site.

\section*{Acknowledgments}

The Hyper Suprime-Cam (HSC) collaboration includes the astronomical communities of Japan and Taiwan, and Princeton University.  The HSC instrumentation and software were developed by the National Astronomical Observatory of Japan (NAOJ), the Kavli Institute for the Physics and Mathematics of the Universe (Kavli IPMU), the University of Tokyo, the High Energy Accelerator Research Organization (KEK), the Academia Sinica Institute for Astronomy and Astrophysics in Taiwan (ASIAA), and Princeton University.  Funding was contributed by the FIRST program from Japanese Cabinet Office, the Ministry of Education, Culture, Sports, Science and Technology (MEXT), the Japan Society for the Promotion of Science (JSPS),  Japan Science and Technology Agency  (JST),  the Toray Science  Foundation, NAOJ, Kavli IPMU, KEK, ASIAA,  and Princeton University.

The Pan-STARRS1 Surveys (PS1) have been made possible through contributions of the Institute for Astronomy, the University of Hawaii, the Pan-STARRS Project Office, the Max-Planck Society and its participating institutes, the Max Planck Institute for Astronomy, Heidelberg and the Max Planck Institute for Extraterrestrial Physics, Garching, The Johns Hopkins University, Durham University, the University of Edinburgh, Queen's University Belfast, the Harvard-Smithsonian Center for Astrophysics, the Las Cumbres Observatory Global Telescope Network Incorporated, the National Central University of Taiwan, the Space Telescope Science Institute, the National Aeronautics and Space Administration under Grant No. NNX08AR22G issued through the Planetary Science Division of the NASA Science Mission Directorate, the National Science Foundation under Grant No. AST-1238877, the University of Maryland, and Eotvos Lorand University (ELTE).

This paper makes use of software developed for the Large Synoptic Survey Telescope. We thank the LSST Project for making their code available as free software at http://dm.lsst.org.

This paper is based on data collected at the Subaru Telescope, which is operated by National Astronomical Observatory of Japan.

We wish to recognize and acknowledge the very significant cultural role and reverence that the summit of Maunakea has always had within the indigenous Hawaiian community. We are most fortunate to have the opportunity to conduct observations from this mountain.

This work is supported in part by MEXT KAKENHI Grant Number 15H05887, 15H05892, 15H05893, and 15K21733.
MT acknowledges supported by JSPS KAKENHI Grant Number JP15K17617. G.H. and Y.L. acknowledge support by the NASA ADAP grant number NNX16AF29G.



\end{document}